# Albert Einstein and Scientific Theology

Max L.E. Andrews[*]

*Department of Philosophy, Liberty University*

*Lynchburg, VA 24551*

18 May 2012

**Abstract**

In recent centuries the world has become increasingly dominated by empirical evidence and theoretic science in developing worldviews.  Advances in science have dictated Roman Catholic doctrine such as the acceptance of Darwinian evolution and Big Bang cosmology.  Albert Einstein (1879-1955) created an indelible impact on the relationship between science and religion.  The question is whether or not his work was deleterious for church doctrine or whether it was compatible with, or even advanced, church dogma.  It's my contention that Einstein revived the relationship between science and theology and did not create a bifurcation between the two. Despite his personal religious beliefs, his work has helped to reinforce the harmonious conjunction of science with religion, which cannot be ignored by succeeding scientists and theologians.

---

[*] mlandrews@liberty.edu

## I. Introduction: Einstein on Theology

Einstein was not an enemy of religion. In 1940 A.S. Yahuda, a professor at Yale, showed Einstein some of Isaac Newton's manuscripts to which Einstein responded with great delight in being able to examine Newton's "spiritual workshop."[1] Einstein was friendly to "God talk."[2] Though he was friendly to religion many historians and biographers deny that his use of the word "God" had any religious implications.[3]

Historically, modern science developed in a religious medieval Europe in an attempt to discover God's creative work. [Modern] science could only arise in a culture dominated by belief in a conscious, rational, and all-powerful Creator. Thus, it could be said that the rise of science required an Eleventh Commandment: "Know thou my handiwork."[4] The universe of space and time is the means by which God has revealed himself to man, as it comes to view under human inquiry to develop and formulate knowledge of God.[5] Science and religion should not be understood as a bifurcation but as a harmonious conjunction. Despite Einstein's lack of religious motivation he had a notable contribution to the advancement of not only science but theology as well.

## II. The Development of Special Relativity

One of the most dramatic changes in recent scientific culture has been Einstein's theory of relativity, which called into question some of the settled assumptions of Isaac Newton (1643-1727). Yet many theologians have been unwise to assume that Newton's ideas were permanent features of the intellectual landscape having based their theology on his model of the universe.[6] For Isaac Newton, space and time, what he linked to the divine *sensorium*, formed a vast envelope that contained all that goes on in the universe. Space and time was independent of all that it embraced and in that sense absolute. Space and time was isomorphic and together with the particle theory of nature formed a mechanistic universe and static concepts that go along with it.[7] Relativity theory must not be confused with its sociological use. On the contrary, relativity refers to an objective relatedness in the universe invariant to any and every observer and for that very reason it necessarily *relativises* the observer's representations of it.[8]

In order to set Einstein's work in proper perspective, in 1865 James Clerk Maxwell had unified electricity and magnetism by developing his equations of electromagnetism.[9] It was soon realized that these equations supported wave-like solutions in a region free of electrical charges or currents, otherwise known as vacuums.[10] Later experiments identified light as having electromagnetic properties and Maxwell's equations predicted that light waves should propagate at a finite speed *c* (about 300,000 km/s). With his Newtonian ideas of absolute space and time firmly entrenched, most physicist thought that this speed was correct only *in one special frame*, absolute rest, and it was thought that electromagnetic waves were supported by an unseen medium called the *ether*, which is at rest in this frame.[11]

Einstein noticed how the Doppler Effect could be applied to electromagnetism.[12] His rather brief paper on the relation between the energy and the mass of an object gave rise to his famous equation $E=mc^2$.[13] This meant that mass energy is proportional to mass. Twice as much mass means twice as much mass energy, and no mass means zero mass energy. The square of the speed of light is called the constant of proportionality. It does the job of converting from the unit in which mass is expressed to the unit in which energy is expressed.[14] With this, Einstein's Special Theory of Relativity (STR) was born.

**III. Einstein and the Epistemic Method**

Lorenzo Valla (1406-1457) developed the interrogative (*interrogatio*) rather than the problematic (*quaestio*) form of inquiry. Valla's mode of inquiry was one in which questions yield results that are entirely new, giving rise to knowledge that cannot be derived by an inferential process from what was already known. This method was similar to the works of Stoic lawyers and educators like Cicero and Quintilian; that is, questioning witnesses, investigating documents and states of affairs without any prior conception of what the truth might be. Valla transitioned from not only using this method for historical knowledge but also applied it as "logic for scientific discovery."[15]

Valla's logic for scientific discovery was the art of finding out things rather than merely the art of drawing distinctions and connecting them together. He called for an active inquiry (*activa inquisitio*). John Calvin (1509-1564) applied this method to the interpretation of Scripture and thus became the father of modern biblical *exegesis* and interpretation.[16] Francis Bacon (1561-1626) applied it to the interpretation of the books of nature, as well as to the books of God, and became the father of modern empirical science.[17]

      This methodology created a split between subject and object, knowing and being, and gave rise to phenomenalism. Newton claimed that he invented no hypotheses but deduced them from observations produced rationalistic positivism, which engulfed contemporary European science. This split's gulf was widened by David Hume's (1711-1776) criticism of causality, depriving science of any valid foundation in necessary connections obtaining between actual events and of leaving it with nothing more reliable than habits of mind rooted in association.[18] Hume weighed heavy in Immanuel Kant's (1724-1804) philosophical development. Given the Newtonian understanding of space and time, Kant transferred absolute space and time from the divine *sensorium* to the mind of man (the transfer of the inertial system), thus intellect does not draw its laws out of nature but imposes its laws upon nature. According to Kant one cannot know the *Ding an Sich* (thing itself) by pure reason; one is therefore limited to the sensual and shaping mental categories of the mind. That which comes through sensation the intuitions are shaped by the mind's *a priori* categories. It is in this sense that Kant played an essential part in the development of the idea that man is himself the creator of the scientific world.

      Throughout Einstein's work, the mechanistic universe proved unsatisfactory. This was made evident after the discovery of the electromagnetic field and the failure of Newtonian physics to account for it in mechanistic concepts. Then came the discovery of four-dimensional geometry and with it the realization that the geometrical structures of Newtonian physics could not be detached from changes in space and time with which field theory operated. Einstein stepped back into stride with Newton and his cognitive instrument of *free invention*. It was *free* in the sense that conclusions were not reached under logical control from fixed premises, and it was *invented* under the pressure of the nature of the universe upon the intuitive apprehension of it. Einstein used Newton and Maxwell's partial differential equations in field theory to develop a mode of rationality called *mathematical invariance*. Mathematical invariance established a genuine ontology in which the subject grips with objective structures and intrinsic intelligibility of the universe.[19]

Einstein's *categories* are not some form of Kantian *a priori* but conceptions that are freely invented and are to be judged by their usefulness, their ability to advance the intelligibility of the world, which is dependent of the observer. As he sees it, the difference between his own thinking and Kant's is on just this point: Einstein understands the categories as *free inventions* rather than as *unalterable* (conditioned by the nature of the understanding). Einstein asserts that *the real* in physics is to be taken as a type of program, to which one is not forced to cling *a priori*.[20]

Principles of method are closely related to empirical observations. As Einstein put it, "the scientist has to worm these general principles out of nature by perceiving in comprehensive complexes of empirical facts certain general features which permit of precise formulation."[21] These principles, not "isolated general laws abstracted from experience" or "separate results from empirical research," provide the basis of deductive reasoning.[22]

There is a long tradition within Christian theology of drawing on intellectual resources outside the Christian tradition as a means of developing a theological vision. This approach is often referred to by the Latin phrase *ancilla theologiae* (a 'handmaid of theology'). The evolution of thought and method from Newton to Einstein vitalized *scientific theology*. Scientific theology argues that the working methods and assumptions of the natural sciences represent the best—or the *natural*—dialogue partner for Christian theology.[23]

Here too logico-deductive argumentation from static concepts and mechanistic systems are rejected. There is another reorientation of man's knowledge leaving epistemic and cosmological dualism behind in operations that have to do with the unity of form and being. Scientific theology is concerned with the discovery of appropriate modes of rationality or cognitive instruments with which to enter into the heart of religious experience, and therefore with the development of axiomatic concepts with which to allow interior principles to be disclosed, and in that light to understand the rational structure of the whole field of God's interaction with man and the world.[24]

Scientific theology takes Einstein's *knowing and being* and his understanding of reality as a whole and applies this method of theology in Christian theology. If the world is indeed the creation of God, then there is an ontological ground for a theological engagement with the natural sciences. It is not an arbitrary engagement, which regresses back to Newtonian engagement, but it is a *natural* dialogue, grounded in the fundamental belief that the God about whom Christian theology speaks is the same God who created the world that the natural sciences investigate.[25]

**IV. Physics, Transcendence, and Immanence**

The Enlightenment restricted knowledge to experience and the phenomenal. Post-Enlightenment thought sought to progress in knowledge while considering the advances the Enlightenment had made. The Christian faith attempted to develop a new relationship between transcendence and immanence. Transcendence has to do with God's being self-sufficient and beyond or above the universe. Immanence corresponds with God being present and active in creation, intimately involved in human history. Newtonian physics did not permit God to be immanent in the universe. This came into question was brought into light by the unmistakable success of science.[26]

Einstein's GTR permitted the possibility that God interacts with the created order without interrupting the physical cause and effect system.[27] Because of Einstein's relativity the Newtonian and Laplacian models have been abandoned. The present discussion of how God interacts with the world has shifted to quantum mechanics. There are over a dozen interpretations, which mathematically describe the quantum world. Objections from the principle of conservation are moot in an Einsteinian universe because it is not causally closed. Even so, certain quantum interpretations reject the principle of conservation such as the Ghirardi-Rimini-Weber interpretation. In a theistic context, Ghirardi-Rimini-Weber makes sense of external causes having an ontological link, the *mass density simpliciter*, to the physical world without violating conservation.[28] Einstein was at odds with Niels Bohr, the father of quantum mechanics, when Bohr suggested the notion of indeterminism on the quantum level. This appalled Einstein, which brought the well-known response of, "God does not play with dice." It is reported that Bohr's response to Einstein was, "Don't tell God what to do."

The most important task for scientific theologians was how to avoid de facto deism—not merely by calling it unorthodox and expressing a dislike for the Newtonian theistic system, but by actually showing why it is an unnecessary conclusion drawn from science. Christian theologians must be in the position to say what they *mean* by God's activity in the world and *how* God's activity can be consistent with the belief that God has created a finite order with a goodness and perfection of its own.[29]

**V. Conclusion**

Albert Einstein's name will be "the watershed of physics" from where all scientific theory will come through the lens of his work. He turned the Newtonian universe upside down and stretched it out with STR and GTR, though he kept the subject's relationship *to* the universe objective as Newton did. Einstein's influence on natural theology has played revitalizing role by providing more mathematical and physical data supporting the metaphysical conclusion that God exists. Albert Einstein's indelible impact has not divorced science from religion; rather, it seems to have done precisely the opposite. Despite Einstein's personal religious beliefs, his work has helped to reinforce the harmonious conjunction of science with religion, which cannot be ignored by succeeding scientists and theologians.

---

[1] Frank E. Manuel, *The Religion of Isaac Newton* (Oxford: Clarendon Press, 1974), 27.

[2] Edward Regis, *Who Got Einstein's Office: Eccentricity and Genius at the Institute for Advanced Study* (Reading, MA: Addison-Wesley, 1987), 24.

[3] Ronald Clark, *Einstein: The Life and Times* (New York: World, 1971), 18-19.

[4] Rodney Stark, *For the Glory of God: How Monotheism Led to Reformations, Science, Witch-Hunts, and the End of Slavery* (Princeton, NJ: Princeton University Press, 2003), 197.

[5] Thomas F. Torrance, "Einstein and Scientific Theology," *Religious Studies* 8 no. 3 (1972): 233.

[6] Alister E. McGrath, *The Science of God: An Introduction to Scientific Theology* (Grand Rapids, MI: Eerdmans, 2004), 29.

[7] Torrance, 236-239.

[8] Ibid., 243.

[9] At this time in 1905 Einstein published a series of articles. These articles included a parallel of Max Planck's work on black body radiation, his PhD thesis which showed how to calculate the size of molecules and work out the number of molecules in a given mass of material based on

their motion, an article on Robert Brown's motion (1827), his article on relativity replacing Newton's laws, and his article on the equivalence of mass and energy. These five works are referred to as Einstein's *annus mirabilis*. Jonathan Allday, *Quantum Reality: Theory and Philosophy* (Boca Raton, FL: CRC Press, 2009), 273.

[10] Vasant Natarajan and Dipitman Sen, "The Special Theory of Relativity," *Resonance* (April 2005): 32.

[11] Ibid., 32 -33.

[12] $v'/v = 1 \pm v/c$ Depending on the relative direction of the light waves and the frame of the observer (*v* signifies measured velocity).

[13] It will be interesting to see how Einstein applies his *invariance* in his work. His argument developed as follows. Let an object in a rest frame simultaneously emit two light waves with the same energy $E/2$ in opposite directions (now having equal but opposite momenta), the object remains at rest, but its energy decreases by $E$. By the Doppler effect, in another frame, which is moving at the velocity v in one of those directions, the object will appear to lose energy equal to $\frac{E}{2}\sqrt{\frac{1-v/c}{1+v/c}} + \frac{E}{2}\sqrt{\frac{1+v/c}{1-v/c}} = \frac{E}{\sqrt{1-v^2/c^2}}$. The difference in energy loss as viewed from the two frames must therefore appear as a difference in kinetic energy seen by the second observing frame. Hence, if *v/c* is very small, in the second frame (the one in motion) the object loses an amount of kinetic energy given by $\frac{E}{\sqrt{1-v^2/c^2}} - E \cong \frac{1}{2} \times \frac{E}{c^2} \times v^2$. Since the kinetic energy of an object with mass *M* moving with speed *v* is given by $(1/2)Mv^2$ (for $v/c \ll 1$), this means that the object has lost an amount of mass given by $E/c^2$. In other words, a loss of energy of *E* is equivalent to the loss in mass of $E/c^2$. This implies equivalence between the mass and energy content of any object. It turns out that for a particle of mass *M*, this quantity is equal to $M^2c^2$. After implementing the Lorentz invariant (and if the frame in which the particle has zero momentum), then the equation $E=mc^2$ is recovered. Ibid., 41-42.

[14] Kenneth William Ford, *The Quantum World: Quantum Physics for Everyone* (Cambridge, MA: Harvard University Press, 2004), 20.

[15] Torrance, 236-237.

[16] Valla served in conjunction with Andrea Alciati (1492-1550) as Calvin's primary influence for his biblical interpretation.

[17] Ibid., 237.

[18] Ibid., 240.

[19] Ibid., 241-242.

[20] Donna Teevan, "Albert Einstein and Bernard Lonergan on Empirical Method," *Zygon* 37 no. 4 (2002): 875-876.

[21] Albert Einstein, *Ideas and Opinions*, Trans. and rev. Sonja Bargmann (New York: Three Rivers, 1982), 221.

[22] Teevan, "Einstein," 877.

[23] McGrath, *The Science of God*, 18-19. There are five distinct classes of things—time, space, matter, energy, and the things relating to conscious life—form with their combinations the known universe. The fifth class must, like the previous, be permanent in quantity, variable in form, and cannot be destroyed. This may be simply labeled as "spirit." In natural science dialogues, this element is often referred to as "God," though it does not necessarily carry the theological meanings with it. This, perhaps, is the sense in which Einstein meant the term "God." T. Proctor Hall, "Scientific Theology," *Monist* 23 (1913): 95.

[24] Torrance, 244.

[25] Both the natural sciences and Christian theology are to engage with the nature of reality—not deciding this in advance, but exploring and establishing it through a process of discovery and encounter. McGrath, *The Science of God*, 21-22.

[26] Clayton Philip, *God and Contemporary Science* (Edinburgh, Scotland: Edinburgh University Press, 1997), 188.

[27] See Thomas Torrance, *Space, Time, and Incarnation* (Edinburgh, Scotland: T&T Clark, 1969).

[28] See Bradley Monton, "The Problem of Ontology for Spontaneous Collapse Theories," *Studies in History and Philosophy of Modern Physics* (2004): 9-10.

[29] Philip, 192.